# IMPROVED BASS MODEL USING SALES PROPORTIONAL AVERAGE FOR ONE CONDITION OF MONO PEAK CURVES

**AHMAD ABU SLEEM**[1,2], **MOHAMMED ALROMEMA**[1,2], **MOHAMMAD A. M. ABDEL-AAL**[1,2*]

[1]Industrial & Systems Engineering Department
King Fahd University of Petroleum and Minerals, Dhahran 31261, Saudi Arabia
[2]Interdisciplinary Research Center of Smart Mobility & Logistics,
King Fahd University of Petroleum and Minerals, Dhahran 31261, Saudi Arabia
Emails: ahmad_yousef_abu_sleem@yahoo.com; mohammadalromaima1992@hotmail.com;
mabdelaal@kfupm.edu.sa; m.abdelaal82@gmail.com
ORCID: 0000-0002-0919-9699 (M.A.M. Abdel-Aal)

## ABSTRACT

This study provides a modified Bass model to deal with trend curves for basic issues of relevance to individuals from all over the world, for which we collected 16 data sets from 2004 to 2022 and that are available on Google servers as "google trends". It was discovered that the Bass model did not forecast well for curves that have a mono peak with a sharp decrease to some level then have semi-stable with small decrement sales for a long time, thus a new parameter based on r1 and r2 (ratios of average sales) was introduced, which improved the model's prediction ability and provided better results. The model was also applied to a data set taken from the Kaggle website about a subscriber digital product offering for financial services that includes newsletters, webinars, and investment recommendations. The data contain 508932 data points about the products sold during 2016-2022. Compared to the traditional Bass model, the modified model showed better results in dealing with this condition, as the expected curve shape was closer to real sales, and the sum of squares error (SSE) value was reduced to a ratio ranging between (36.35-79.3%). Therefore, the improved model can be relied upon in these conditions.

## 1  INTRODUCTION:

Frank Bass created the Bass model, sometimes known as the Bass diffusion model. It is made up of a simple differential equation that explains how new items get accepted in a community. The model rationalizes how existing and future adopters of a unique product interacts.

(Dodds, 1973) introduced the Bass model framework, which is used to create a long-term forecast of cable television uptake. (Bass et al., 1994) extended the original model by incorporating choice variables such as pricing and advertising. (Golder et al., 1998) presented a basic model for the growth of new consumer durables that focused on affordability rather than diffusion. (Krishnan et al., 1999) used a modified version of (Bass et al., 1994) generalized Bass model (known as GBM) to build optimum pricing strategies that are compatible with empirical data.

By taking into account the trend of "extra-Bass" skew in data or basic population variation, (Lee, 2002) presents an approach for explaining variances in the estimated parameters of the Bass model, including the value of the coefficient of invention (p), coefficient of imitation (q), and market penetration rate (c). Furthermore, (Bass, 2004) mentions that in connection with the journal's fifty anniversary, the "Bass Model" publication was recognized as one of the Top 10 Most Significant Papers published in Management Science's fifty-year history, which was a heartening recognition of the model's impact on the field of management science.

(Sood et al., 2009) examined the sales penetration of 760 sorts drawn from 21 items and 70 countries in their study. They also assessed the efficiency of functional data analysis (FDA), a nonparametric methodology that has yielded impressive results in the statistics community, in forecasting market penetration. For projecting eight features of market penetration, the researchers compared FDA's predictive ability to that of various other models, including the traditional Bass model, Expected Means, Last Examination Projection, a Meta-Bass model, and a Reinforced Meta-Bass model.

(Jiang et al., 2012) develop a Generalized Norton-Bass (GNB) model that separates the two types of substitutions. (Lee et al., 2014) propose a novel method. It is feasible to anticipate new product demand ahead of time using the Bass model and statistical and machine learning approaches. (Fan et. al., 2017) developed a new technique for product sales forecasting that integrates the Bass/Norton model with sentiment analysis to incorporate historical sales data and online review data.

(Liang, 2012) employ the BASS model, ARIMA model, and GM model to calculate the life cycle in order to support the healthy and speedy growth of China's dairy business. (Wang et al., 2014) describe a hybrid particle swarm optimization (HPSO) approach for improving generalized Bass diffusion model parameter estimations. To address the shortcomings of the current Bass model (Wang et al., 2017), propose the westernization solution of differential equations.

(Zhang et al., 2018) investigate the possibilities for power substitution in several industries in Japan and South Korea between 2020 and 2050. To replicate the model, (Zhao et al., 2018) use China's wind power sector as an example. (Sweet et al., 2019) show that using the Bass Model enhanced students' ability to reflect on practice as indicated by their writing, and that it is a post-practicum approach with the potential to boost learning.

In order to sell product information, microblogs are used. Prior studies concentrated on the attributes of information content, information sources, and knowledge while ignoring the influence of user retweeting decisions on information popularity. (Han et al., 2022) create a two-phase framework for product data dissemination in their work by examining the decision-making methods of both sets of users when they retweet content from various information sources. The exponential- and power-function augmented methods are suggested as two upgraded Bass approaches using this model to examine user interest dropoff percentages. According to the outcomes of the experiments and model contrasts, the exponential-function better model performs greater than the Bass, Gompertz, and power-function better models and is suitable for predicting the popularity of a product's information prior to its debut.

Despite the fact that (Harald et al., 2023) study the simple and ideal scenarios; the answers could make clear key strategic elements that would otherwise be hidden by complication in the real world, such as the causes of a market's earliest expansion's sluggishness. Their research also looks at how existing models might be applied to evaluate how business practices and market trends in the digital economy have changed.

(Zhang et al., 2022) offer a method based on social media data for projecting the box office of upcoming movies. First, quantities of movie-related web reviews are gathered from microblogs. Second, various key aspects, such as the emotional worth of microblog texts. Two

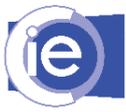 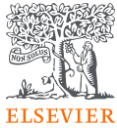 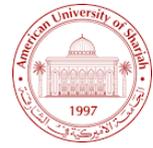

CIE50 Proceedings, October 30 – November 2, 2023

American University of Sharjah, UAEenhanced Bass models are produced for projecting movie box office in the first week and over the entire diffusion cycle, respectively, using the characteristics obtained to expand the parameters of the Bass model. By continuously fitting data, they finally finish the forecasting assignment. The findings from experiments suggest that the modified Bass technique proposed in their research outperforms standard forecast models and prior studies in terms of overall performance.

(Xiaoyu Li et al., 2020) they analyze the changing law of consumer quantity from the macro level in which may enable E-commerce platforms appropriately estimate it. To begin, they highlight the distinct aspect of B2C commerce platforms in comparison to traditional goods or technology, namely indirect network externality. And they use this characteristic, as well as the elements that drive consumers' and businesses' adoption of B2C trade, to create an expanded Bass Model. Finally, they validate their model using data from Chinese online buyers.

Another improved model for fashion clothing from (Xiaoxi Zhou et al., 2020) research. In order to address the issue of a lack of sales data for fashion, the study proposes a scientific, quantitative method for projecting new clothing products using historical sales data of comparable clothes. The improved Bass model incorporates consumer preferences and seasonality more fully, improving prediction precision for fast fashion clothes demand.

## 2   PROBLEM STATEMENT

This study presents an enhanced version of the Bass model to address the challenges of forecasting trend curves for global issues relevant to individuals worldwide. To achieve this, we gathered 16 datasets from 2004 to 2022, sourced from "Google Trends" available on Google servers. Our analysis revealed that the traditional Bass model exhibited limitations in accurately predicting curves characterized by a single peak followed by a sharp decline, transitioning into a phase of relatively stable, but gradually decreasing sales over an extended period. To overcome this limitation, we introduced a novel parameter based on the ratios of average sales (r1 and r2), resulting in improved predictive capabilities and more accurate forecasts.

Furthermore, we applied the modified model to a dataset obtained from the Kaggle website. The dataset pertains to a subscription-based digital product offering financial services, including newsletters, webinars, and investment recommendations. The dataset comprises 508,932 data points representing product sales between 2016 and 2022.

By leveraging these datasets and introducing the enhanced Bass model, we aim to provide more reliable predictions and insights into mono peak trend curves followed by stable sales period.

## 3   METHODOLOGY

The first step is data collection as appears in the Figure (1), where a set of historical databases were collected from Google Trend about the percentage of search times on a specific interest, in addition to obtaining a historical database from the Kaggle website related to subscribing to financial services for several services products over the years 2016-2021.

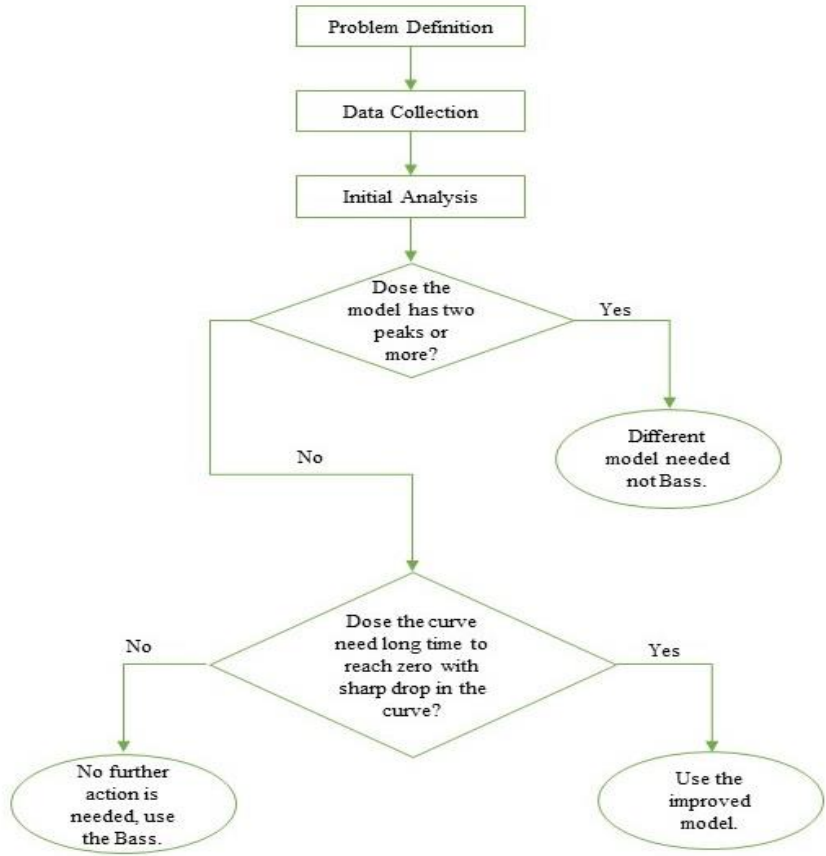

Figure 1: Methodology flow chart

Symbol table contains the Variable/Parameter for the research equations in Table (1):

Table 1: Variable explanation summary

| Variable/Parameter | Meaning |
|---|---|
| $a$ | Linear regression model constant number. |
| $b$ | The factor of the linear variable in the linear regression model. |
| $c$ | The factor of the quadratic variable in the linear regression model. |
| $d(t)$ | The predicted demand of time $t$ by the Bass model. |
| $D(t-1)$ | The cumulative demand by time $t-1$. |
| $r1, r2$ | Ratios of average sales. |
| $d_m$ | The actual demand of time $t$. |

Then we created an initial analysis in which the curves were classified into curves in which the bass works well and curves in which the bass does not work well due to some characteristics of these curves such as curves that have a mono peak with a sharp decrease to some level then have semi-stable with small decrement sales for a long time that required further improvements on the Classical model to deal with this issue.

We have noticed the inability of the Bass model to deal with these mentioned curves, as the Bass model was have high SSE in this case so it was necessary to amend equation number (1) to become as it is in the form of equation number (2) and (3), where a new part has been added to the model that collects the average relative sales each month for all readings to give a better curve shape, as it can be relied upon to reduce SSE and obtain better and more accurate forecasts, especially over long periods of time.

Bass model equation:

$$d(t) = a + bD(t-1) + cD(t-1)^2 \qquad (1)$$

Modified Bass model equations:

$$d(t) = a + bD(t-1) + cD(t-1)^2 + \frac{r1 \sum_{i=0}^{n} d_m}{n} \qquad (2)$$

$$d(t) = a + bD(t-1) + cD(t-1)^2 - \frac{r2 \sum_{i=0}^{n} d_m}{n} \qquad (3)$$

Considering a reference point where the tail starts from 50% of the total height of the curve, the reference values for r1 and r2 are set as follows: ref(r1) = 0 and ref(r2) = 0.5. In this context, $d_m$ represents the actual monthly demand from a specific data point.

The equations for calculate r1,r2:

$$r1 = \text{ref}(r1) + (\text{tail per} - 0.5) \times 1.6 \qquad (4)$$

$$r2 = \text{ref}(r2) - (\text{tail per} - 0.5) \times 1.6 \qquad (5)$$

The term 'tail per' represents the ratio of the length of the tail to the overall length of the shape.

## 4   RESULTS & DISCUSSION

Here are some examples of curves where some of them work appropriately with the Bass model whereas the others requested further adjustment so we use the modified model for them. The first example is about the trend "Kingston ram DDR2" from 2006-2022 for this curve Bass model was a sufficient and good predictor for it with an SSE of 13,022.69 as appears in the Figure (2) and Figure(3).

The second example is about the trend "Mercedes E class 2015" from 2014-2022 for this curve Bass model did not work well with bad SSE of 22,061.11 as appears in the Figure (4). Whereas, when using the improved model the SSE value reduced to 14,041.68 it is equal to 36.35% enhancement of the model prediction as appear in Figure (5).

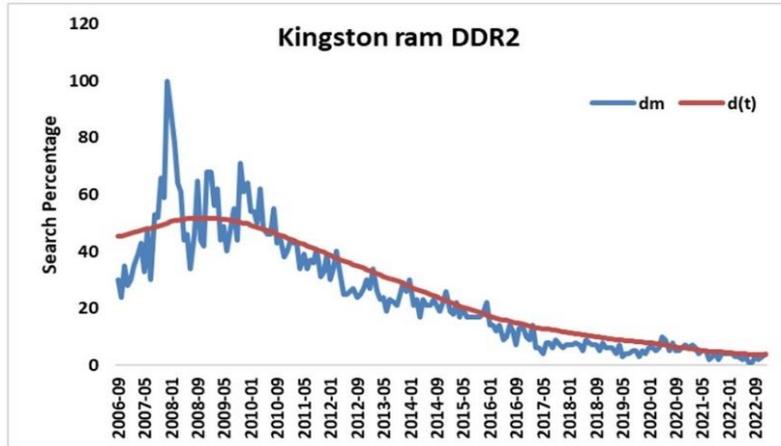

Figure 2: Kingston ram DDR2

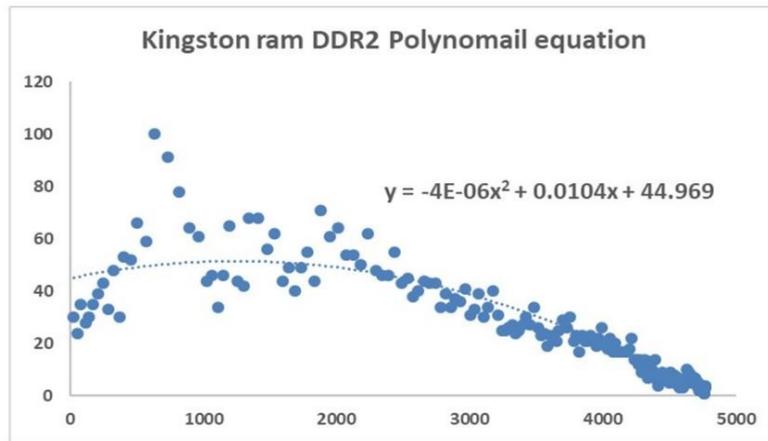

Figure 3: Kingston ram DDR2 Polynomial equation

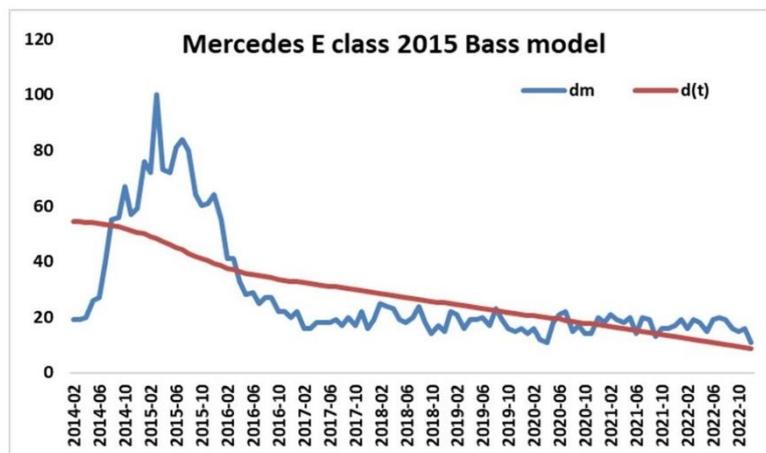

Figure 4: Mercedes E class 2015 Bass model

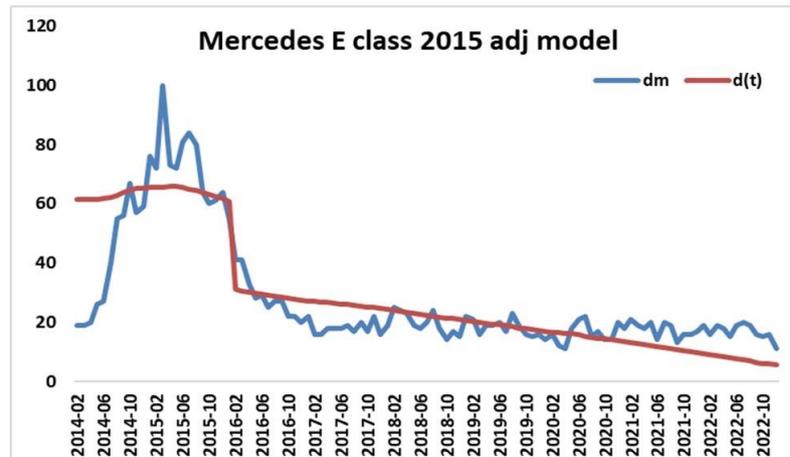

Figure 5: Mercedes E class 2015 adjusted model

The third example is about the trend "VIBER" from 2010-2022 for this curve Bass model did not work well with bad SSE of 179,525.74 as appears in the Figure (6). Whereas, when using the improved model the SSE value reduced to 37,197.51 it is equal to 79.3% enhancement of the model prediction as appear in Figure (7).

The shape of the prediction curve when using the improved model is much better than the shape of the curve that resulting from the classical Bass model, as clearly appeared in the mentioned figures.

The results showed the ability of the improved model to deal with this type of curve, and therefore it can be relied upon if the behavior of users towards a specific product adopts the same current behavior for this condition.

The last example is for the finance services company data for years 2020-2022 for the four quarters, for this curve Bass model work well with SSE of 301,500,362.4 as appears in the Figure (8). Whereas, when using the improved model the SSE value reduced to 180,145,992.2 that means further improvement in the results, it is equal to 40.25% enhancement of the model prediction as appear in Figure (9).

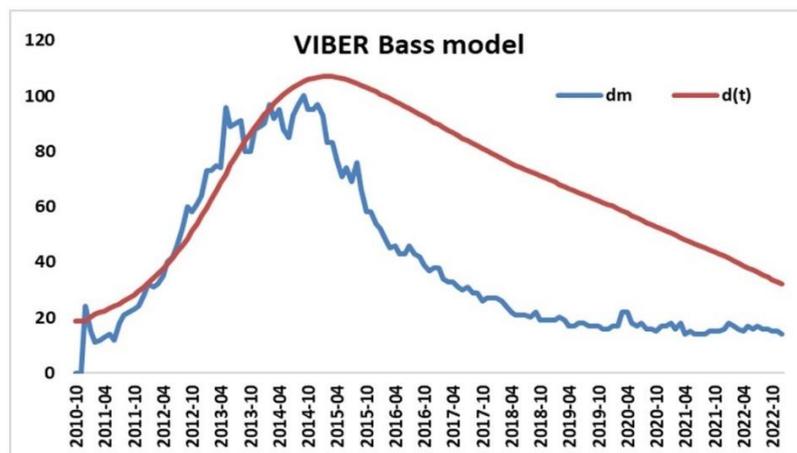

Figure 6: VIBER Bass model

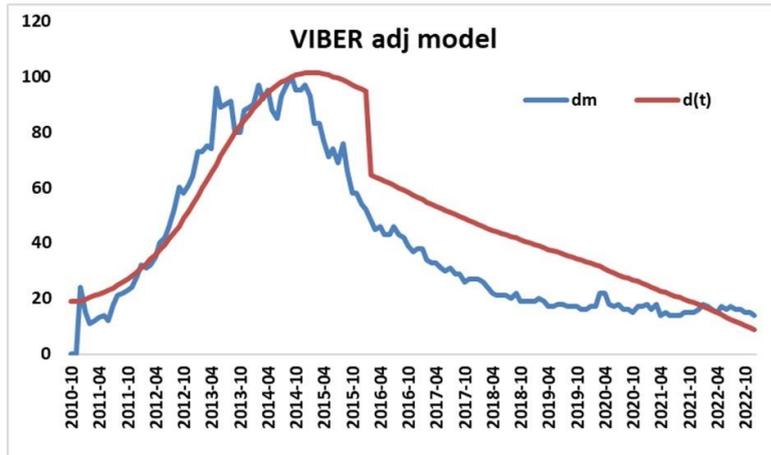

Figure 7: VIBER adjusted model

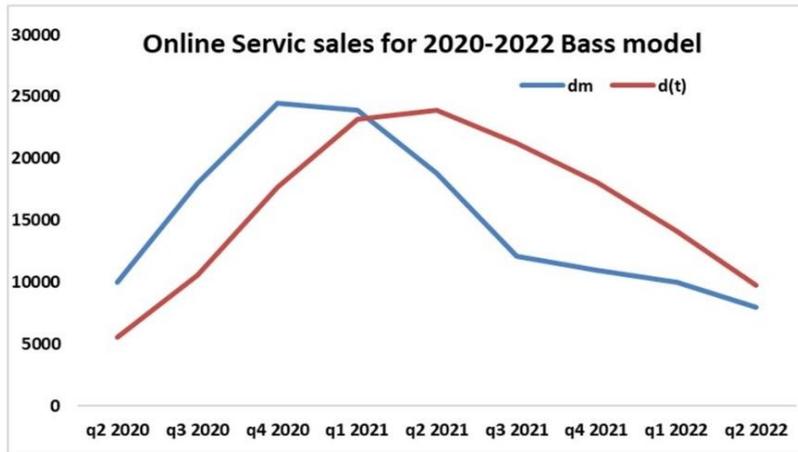

Figure 8: Online service sales Bass model

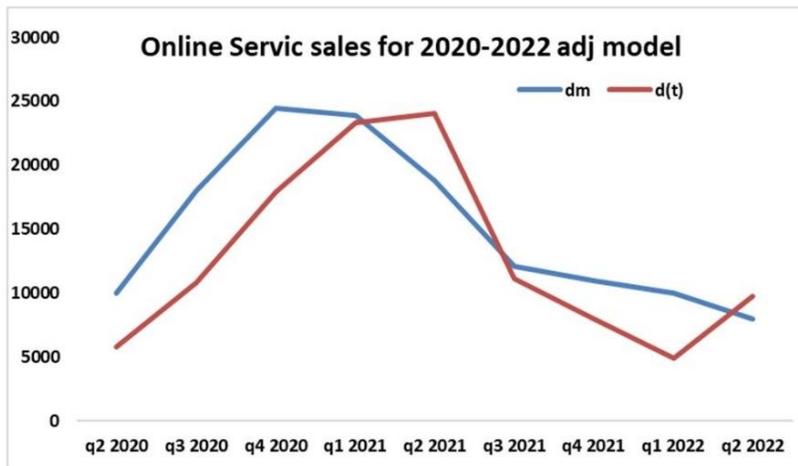

Figure 9: Online service sales adjusted model

## 5 CONCLUSION

It has been observed that the Bass model has limitations in accurately forecasting curves that exhibit a single peak followed by a sharp decrease, eventually stabilizing with a gradual decline in sales over an extended period. To address this issue, a new parameter based on the ratios of average sales (r1 and r2) has been introduced. This parameter enhancement has significantly improved the prediction ability of the Bass model, leading to better results in such scenarios.

The enhanced model demonstrated superior performance compared to the conventional Bass model when handling this specific condition. The modified model achieved a closer alignment between the projected curve shape and actual sales data, resulting in a substantial reduction in the SSE by a range of 36.35 to 79.3%. Consequently, the improved model can be considered a reliable choice for addressing this condition.

## 6 FUTURE WORK

Future work will involve analysing two, three, and multi-peak curves and developing a model specifically tailored to handle these complex cases.

## 7 ACKNOWLEDGMENT

This research was supported by the Interdisciplinary Research Centre of Smart Mobility & Logistics.